\documentclass[acmlarge]{acmart}
\usepackage[T1]{fontenc}
\usepackage{booktabs}
\usepackage{array}
\usepackage{hyperref}
\usepackage{enumitem}
\usepackage{xcolor}
\usepackage{natbib}
\usepackage{longtable}
\usepackage{xltabular}
\usepackage{ifthen}

\keepXColumns
\newcommand{\secondedit}[1]{#1}

\newcommand{\minoredit}[1]{#1}
\title{From Neglect to Resignation: Understanding the Evolution of Mobile Privacy}
\author{Yibo Meng}
\orcid{0009-0009-0370-5456}
\affiliation{%
  \institution{Cornell University}
  \city{New York}
  \country{USA}
}
\email{yibom2@illinois.edu}

\author{Ziqi Liu}
\orcid{0009-0007-8755-5744}
\affiliation{%
  \institution{University of Wisconsin-Madison}
  \city{Madison}
  \country{USA}
}
\email{ziqil@cs.wisc.edu}

\author{Lionel Z. Wang}
\orcid{0009-0006-5387-8812}
\affiliation{%
  \institution{The Hong Kong Polytechnic University}
  \city{Hong Kong}
  \country{Hong Kong}
}
\email{lionel-z.wang@connect.polyu.hk}

\author{Shuning Zhang}
\orcid{0000-0002-4145-117X}
\affiliation{%
  \institution{Tsinghua University}
  \city{Beijing}
  \country{China}
}
\email{zsn23@mails.tsinghua.edu.cn}

\author{Xiaolan Ding}
\orcid{0009-0006-9956-4560}
\affiliation{%
  \institution{Independent Researcher}
  \city{Zhengzhou}
  \country{China}
}
\email{13838038961@163.com}

\setcopyright{cc}
\setcctype{by}
\acmJournal{IMWUT}
\acmYear{2026}
\acmVolume{10}
\acmNumber{4}
\acmArticle{242}
\acmMonth{12}
\acmDOI{10.1145/3857977}

\AtBeginDocument{%
  \hypersetup{%
    colorlinks=true,
    allcolors=ACMDarkBlue,
    citecolor=black
  }%
}

\begin{document}

\begin{abstract}

This decade-long longitudinal qualitative study follows 31 smartphone users in China across four interview waves (2014–2023) to examine how mobile privacy understanding evolves alongside shifting computing infrastructures. Findings reveal three stages shaped by distinct interactional mechanisms: early functional acceptance of permission-based data access, growing vigilance triggered by behavioral tracking feedback loops, and eventual resignation as AI-driven inference made ecosystem-level profiling feel inescapable. Rather than a stable attitude, digital resignation emerges through cumulative infrastructure exposure, as each new system layer retroactively reinterpreted earlier permissions and exhausted users' capacity for self-management. The study challenges permission-centric design assumptions and proposes inference visibility, context isolation, and automated privacy agency as system-level responses.

\end{abstract}

\maketitle

\section{Introduction}

Over the past decade, mobile ubiquitous computing has been reorganized by three infrastructure shifts: permission-gated access to device resources \citep{felt2012android, wijesekera2015android}, behavioral telemetry across apps and platforms \citep{jin2018why, ullah2022privacy}, and AI-mediated inference from accumulated traces \citep{raij2011privacy, wachter2019right}. Permission systems made privacy legible as a local decision about whether an app could use location, contacts, photos, cameras, microphones, or other resources. Behavioral telemetry turned everyday use into streams of searches, clicks, mobility traces, dwell time, and social interactions that could return as targeted ads, ranked feeds, and cross-platform personalization. AI-mediated inference further loosened the connection between privacy risk and any single disclosure event by enabling services to derive identity-level, emotional, and routine-based interpretations from combinations of traces \citep{asthana2024know}. Mobile privacy is therefore not only a question of concern or disclosure preference. It is a question of how changing sensing and inference infrastructures reshape the conditions under which users notice, explain, and attempt to manage privacy.

Research in HCI and related fields has used concepts such as the privacy calculus \citep{dinev2006extended}, the privacy paradox \citep{barnes2006privacy}, and privacy fatigue \citep{choi2018role} to explain how people evaluate disclosure, continue using systems despite concern, and become exhausted by repeated privacy management. Much of this research examines a particular technology or privacy decision at one point in time, such as responding to a permission prompt, sensor, notice, wearable, or smart environment. This focus makes it difficult to explain how users later revise the meaning of an earlier decision when new forms of tracking and inference become visible. A permission granted for location, contacts, photos, or voice input may initially appear to be functional access for a task; years later, the same permission may be understood as an entry point into behavioral tracking, cross-app personalization, or inference. What remains unclear is how repeated encounters with changing mobile infrastructures alter users' interpretations, gradually produce resignation, and indicate where design intervention is needed.

To address this gap, we conducted a decade-long longitudinal qualitative study of mobile privacy in China from 2014 to 2023. The study followed 31 smartphone users through four waves of in-depth interviews in 2014, 2017, 2020, and 2023, with an 80.6 percent retention rate. China is not treated here as a universal case. Rather, it is an analytically dense site for observing rapid mobile infrastructure transition, including the movement from standalone applications toward super-app ecosystems, cross-context data aggregation, payment and service integration, and increasingly AI-integrated services. This setting allows us to examine how the same users revised their privacy interpretations as everyday mobile interaction became embedded in more consolidated sensing, telemetry, synchronization, and inference systems.

Our findings show that digital resignation emerged through cumulative infrastructure exposure rather than as a passive default, stable trait, or simple loss of concern. We identify three stages. First, participants interpreted direct access through a functional framing: permission prompts translated architectural access into task requirements, making contacts, photos, microphones, cameras, and location appear necessary for immediate app functions \citep{felt2012android}. Second, behavioral feedback loops made tracking perceptible. Targeted advertisements, recommendation repetition, cross-platform content echoes, and system access indicators turned background telemetry into observable responses, prompting vigilance and localized tactics such as obfuscation or account separation \citep{eslami2015always}. Third, AI-mediated inference made ecosystem-level profiling feel difficult to bound. Participants recognized that prompts, voices, locations, platform histories, and other ordinary traces could be aggregated into inferences about who they were, what they did, and what circumstances they faced. Resignation arose as users repeatedly encountered a widening gap between local privacy actions and infrastructure-level inference capacities \citep{shao2022you}.


This trajectory exposes a structural mismatch in mobile privacy design: user concerns increasingly attach to ecosystem-level inference, while operating systems and applications still largely present privacy as user-managed access to discrete resources. The paper makes three contributions to HCI and IMWUT. Empirically, it provides longitudinal evidence that the same users retrospectively reinterpreted earlier permissions as new infrastructure layers became perceptible. \secondedit{Theoretically, it identifies a temporal mechanism---cumulative infrastructure exposure---that explains how later tracking and inference feedback causes users to reinterpret earlier, seemingly functional disclosures as profiling inputs. Existing concepts such as privacy fatigue, cynicism, and surveillance realism describe depletion, mistrust, or normalization after privacy risks and constraints have become salient; they do not specify how users come to recognize those risks over time \citep{choi2018role, hoffmann2016privacy, dencik2017advent}.} For design, it argues that transparency must be paired with system-level agency. \secondedit{Inference visibility, context isolation, and automated privacy agents are instantiated as a ledger that shows which data support an inference, separate data and memory spaces for different contexts, and a policy agent that checks user-defined rules before sharing data across contexts or taking cross-context actions.}

\section{Related Work}

This study treats privacy as a long-term interactional problem in ubiquitous computing systems. We review four bodies of work: mobile and ubiquitous sensing privacy; behavioral tracking and platformized data flows; user adaptation, privacy fatigue, and digital resignation; and longitudinal HCI/UbiComp research. Across these areas, the central gap is temporal: prior work explains privacy expectations, tracking experiences, and coping strategies, but rarely follows the same users as mobile infrastructures move from permission-based access to persistent telemetry and inference.

\subsection{Privacy in Mobile and Ubiquitous Sensing Systems}
Ubiquitous computing has long complicated privacy because sensing is embedded in ordinary devices, environments, and routines. Foundational HCI and UbiComp accounts frame privacy as boundary regulation and situated information flow rather than secrecy alone \citep{altman1975environment, westin1968privacy, palen2003unpacking, dourish2006collective}. Contextual integrity similarly locates privacy problems in transfers of information across contexts that violate situated expectations \citep{nissenbaum2004privacy, barth2006privacy}. This framing is central for mobile systems, where sensors, accounts, applications, cloud services, and nearby devices make information movement part of everyday interaction.

Prior UbiComp work demonstrates the breadth of mobile and wearable sensing. Smartphone studies such as StudentLife used phones to infer mental health, academic performance, sleep, sociability, activity, and behavioral trends \citep{wang2014studentlife}, while in-the-wild context recognition shows how multi-modal sensors support activity and context inference outside controlled settings \citep{vaizman2018context}. Wearable cameras, fitness trackers, health wearables, and privacy-aware sensing platforms extend this surface to bystanders, physiological traces, and AI-based processing in daily life \citep{hoyle2014privacy, velykoivanenko2021those, fernandes2024habitsense}. Mobile privacy research then asks how users understand access to location, contacts, photos, microphones, cameras, files, and SMS through permissions. Users often have limited attention to prompts \citep{felt2012android}; their expectations based on app function may diverge from actual collection and use \citep{lin2012expectation, wijesekera2015android}; and they value data types and granularities differently, especially location and richer behavioral interpretations \citep{staiano2014money, fan2024evaluating}.

The permission boundary is incomplete. Apps may collect or infer information through background access, third-party libraries, side channels, covert channels, or platform behaviors outside the decision users remember making \citep{chitkara2017does, reardon201950, shklovski2014leakiness}. Responses include stream-processing models of personal data access \citep{li2017privacystreams} and notices, nudges, labels, personalized notifications, dashboards, and runtime indicators designed to make data practices more visible and actionable \citep{schaub2015design, zhang2016privacy, almuhimedi2015your, kelley2010standardizing, emami2020ask, jackson2018addressing, mhaidli2020listen}. AR privacy research further shows that low-level access control is insufficient when awareness and consent depend on activity and social relation \citep{ohagan2022privacy}. This literature explains sensing, permissions, visibility, and control; its gap is temporal reinterpretation. It rarely shows how the same users later reclassify functional permissions as components of a larger data infrastructure after tracking and inference become perceptible.

\subsection{Behavioral Tracking, Persistent Data Flows, and Platformized Mobile Ecosystems}
As mobile ecosystems matured, data collection expanded beyond discrete resource access to continuous behavioral telemetry from users' interactions, movements, and cross-app activity \citep{jin2018why, reardon201950, ullah2022privacy}. These signals support personalization, targeted advertising, and cross-platform account coordination. Mobile advertising and analytics research makes this infrastructure visible: targeting relies on information flows among apps, ad libraries, analytics firms, identifiers, profiles, and ad-delivery platforms \citep{jin2018why, ullah2022privacy}. App leakage and permission-circumvention studies show that OS prompts do not fully capture these flows, especially when third-party SDKs or side channels move data beyond the visible interaction boundary \citep{reardon201950}. Work on internet mental models likewise shows that users often struggle to identify where data go, who receives them, and how actors participate in collection and sharing \citep{kang2015my}.

Recommendations, feeds, and ads are important because they make tracking experienceable. Users develop folk theories when previous actions reappear as ranked content, targeted advertisements, or altered information environments \citep{eslami2015always, eslami2016first, rader2015understanding, devito2017algorithms, siles2020folk, barbosa2021who}. Persistent data flows also extend beyond the phone. Smart homes, personal assistants, wearable cameras, AR systems, and IoT environments distribute collection across devices, places, households, and bystanders \citep{choe2011living, zheng2018user, lau2018alexa, abdi2021privacy, pierce2019smart, hoyle2014privacy, ohagan2022privacy}. IMWUT security work showing that smart-home network traffic can reveal routines further underscores that sensitive context can be inferred from infrastructural traces rather than explicit disclosure \citep{zou2023iotbeholder}. Together, these studies move privacy analysis from isolated interface decisions toward coordinated sensors, networks, platforms, accounts, and inference services.

Platform ecosystems intensify these dynamics. In China, WeChat has been analyzed as an infrastructure joining messaging, payment, services, mini-programs, and governance \citep{plantin2019wechat}, and super-app/app-store arrangements consolidate social, commercial, and technical functions \citep{jia2022superapps}. Studies of Chinese users show that WeChat's social and professional indispensability can blur public, private, and work boundaries \citep{chen2018privacy}, while valuation of personal information is shaped by platform dependency and perceived data value \citep{tang2021chinese}. China's Personal Information Protection Law, effective in 2021, also changed the formal governance environment \citep{greenleaf2021completed}. The gap is not whether platforms track or consolidate data, but how the same users reinterpret privacy as an ecosystem moves through this platform transition. Existing work usually analyzes platforms, users, or regulation at one time point; it less often captures longitudinal reinterpretation across standalone apps, super-app integration, and AI-enabled inference.

\subsection{User Adaptation, Privacy Fatigue, and Digital Resignation}
Users do not only notice data infrastructures; they adapt to them. Prior work documents coping strategies including obfuscation, selective disclosure, account separation, withholding engagement signals, manipulating recommendation systems, and reducing participation in particular platform contexts \citep{brunton2015obfuscation, eslami2016first, velkova2021algorithmic, karizat2021algorithmic}. Users also learn to ``work with'' algorithmic systems by inferring which inputs matter, how feedback loops operate, and how to avoid undesirable signals \citep{lee2015working, seaver2017algorithms, ytre2021folk}. UbiComp sensing studies show similar utility/privacy tensions: fitness-tracker users value health feedback while holding incomplete mental models of processing and inference \citep{velykoivanenko2021those}; lifelogging camera users may prefer in situ control over capture to burdensome later review \citep{hoyle2014privacy}; and AR bystander privacy depends on activity type and social relationship \citep{ohagan2022privacy}.

These adaptations have limits. Privacy apathy emphasizes the belief that little can be done when protection seems ineffective or socially unrealistic \citep{hargittai2016can}. Privacy fatigue highlights exhaustion and reduced willingness to manage privacy after repeated decisions, breaches, and complex controls \citep{choi2018role, shao2022you}. Privacy cynicism captures mistrust, uncertainty, powerlessness, and resignation toward institutional data practices \citep{hoffmann2016privacy}, while surveillance realism describes unease combined with normalization when alternatives are hard to imagine \citep{dencik2017advent}. Digital resignation is especially relevant because it names a condition in which people want control over what digital entities know about them but feel unable to achieve it, linking resignation to routine corporate practices and power asymmetries rather than indifference \citep{draper2019corporate}. Recent work maps these related concepts as forms of constrained agency under pervasive data collection and inference \citep{draper2024privacy}. \secondedit{These concepts primarily characterize states or responses once privacy risks or limits on choice are already salient. Our study addresses an earlier temporal question: how users come to recognize those risks and arrive at resignation as mobile infrastructures change. Cumulative infrastructure exposure captures this process by showing how later tracking and inference feedback turns earlier, seemingly functional disclosures into recognizable inputs to profiling.}


Inference intensifies the limits of adaptation. Early work on innocuous wearable sensors showed that low-risk streams can reveal sensitive attributes when combined and modeled \citep{raij2011privacy}. More recent work examines concerns about AI-based inferences and the difficulty of governing what systems derive from ordinary data \citep{wachter2019right, asthana2024know}. In mobile and ubiquitous settings, inference can draw on interaction histories, sensor traces, locations, conversations, wearable data, and cross-device synchronization. \secondedit{By following the same users across these changes, we trace how functional acceptance can become vigilance and, when perceived infrastructural capacity outpaces available control, resignation.}

\subsection{Longitudinal HCI and UbiComp Research}
The preceding gaps make a longitudinal design necessary. HCI scholars argue that longitudinal studies reveal how human--technology relationships change after novelty fades and systems become embedded in daily life \citep{karapanos2021advances, kjaerup2021longitudinal}. Reviews of HCI and mobile HCI methods note the value of empirical, field-oriented work while emphasizing the difficulty of sustained access, repeated measurement, retention, and interpretation of change \citep{kjeldskov2012longitudinal}. Privacy research has used longitudinal designs to study disclosure, trust, and trade-offs with experience, but often around a specific app, experiment, or adoption period rather than a decade of infrastructural change \citep{keith2014longitudinal}.

For ubiquitous computing, this methodological issue is substantive: the system being studied does not remain stable. Mobile operating systems add dashboards and indicators; platforms expand tracking and recommendation; accounts synchronize across devices; and AI-mediated inference becomes more salient. Cross-sectional work can capture one configuration, but not how earlier understandings are revised after later capabilities become visible. Prior literature therefore leaves three elements insufficiently combined: the same users followed over time; multiple technological paradigms from permissions to telemetry and inference; and a platformized mobile ecosystem in which these shifts become part of everyday dependency. This paper addresses that gap with longitudinal evidence and a conceptual account of how evolving UbiComp infrastructures reshape perceived control, agency, and adaptation.

\section{Methods}

\subsection{Research Design}

This study adopts a qualitative longitudinal research design to understand how users gradually develop and adjust their understanding of mobile privacy risks over extended periods of smartphone use. Specifically, the study examines how different smartphone interaction mechanisms and hardware--software features influence users' interpretations of privacy, as well as the strategies they employ to manage privacy in everyday mobile computing environments.

This longitudinal study is situated within the rapidly evolving mobile internet ecosystem of China between 2014 and 2023. This specific geographical and temporal context serves as an exceptional natural laboratory for investigating privacy perceptions. During this nearly decadal span, the regional technological landscape transitioned from the initial proliferation of standalone smartphone applications to the dominance of highly integrated super applications. These ubiquitous platforms seamlessly consolidated communication, financial transactions, mobility services, and algorithmic content distribution into unified digital environments. Furthermore, this period witnessed the pervasive establishment of cross platform data synchronization and the eventual implementation of comprehensive national privacy legislation in 2021. Consequently, observing individuals within this specific ecosystem provides a magnified and accelerated view of how people negotiate privacy boundaries when technical systems become deeply infrastructural and unavoidable in everyday life.

Rather than treating privacy concern as a static attitude, this research focuses on how users continuously reinterpret privacy issues through their ongoing interactions with mobile computing systems \citep{palen2003unpacking}. In this perspective, privacy awareness is not assumed to exist prior to technology use; instead, it gradually emerges through concrete experiences with applications, platforms, and device capabilities.

To capture this evolving process, we conducted multiple rounds of follow-up interviews with the same group of users. In total, four rounds of interviews were carried out in 2014, 2017, 2020, and 2023. This longitudinal design allows us to observe how users' understandings of privacy risks change over nearly a decade, alongside the development of mobile computing systems. It also enables us to examine the interactions and tensions between technological change and users' evolving interpretations of privacy in everyday mobile use.

\subsection{Participants}

A total of 31 participants were recruited for this study. Participants first entered the study in 2014 and were invited to continue participating in subsequent interview rounds. By following the same group of individuals across multiple time points, the study adopts a panel-style sampling design that allows us to examine how changes in both technological environments and participants' life trajectories influence their evolving understanding of mobile privacy.

At the time of the first interview in 2014, participants ranged in age from 19 to 54. The sample included individuals from diverse backgrounds in terms of gender, urban--rural residence, educational attainment, and occupation. This diversity enabled the study to capture smartphone use experiences across different levels of digital literacy, socioeconomic positions, and work environments. Detailed participant characteristics are provided in Appendix A.

Participants were initially recruited through a combination of purposive sampling and snowball sampling. At the beginning of the study in 2014, the last author recruited potential participants through the researchers' social networks, community connections, and referrals from existing participants. After completing the first interview, participants were invited to take part in future follow-up interviews. The research team maintained contact with participants through phone calls, instant messaging, or email, and invited them to participate in later interview rounds.

Because the study spans nearly a decade, some participants were unable to continue participating in later stages due to job changes, relocation, or personal circumstances. As a result, the number of participants in each interview round varied slightly over time.

Despite the extended duration of this research, the study achieved a remarkable retention rate of 80.6 percent, with 25 of the original 31 participants completing the final interview round in 2023. This high retention in a decadal longitudinal qualitative study was actively facilitated through deliberate rapport building and sustained ethical engagement. The research team maintained a consistent point of contact, ensuring that individuals interacted with the same primary researcher across multiple waves to foster familiarity and trust. Between formal interview rounds, the team utilized low frequency and non intrusive communication methods to share seasonal greetings and express ongoing appreciation for their contribution to scientific research. This continuous yet respectful engagement strategy, combined with transparent communication regarding how their previous data informed the ongoing study, transformed the individuals from mere subjects into long term collaborative stakeholders in the research process. Inclusion criteria for the initial recruitment included: (1) having basic experience using smartphones, (2) being able to describe their everyday use of mobile applications, and (3) being willing to participate in the interview and possible follow-up interviews in later stages of the study. Exclusion criteria included individuals who had never used smartphones or mobile internet services, individuals who were unable to complete the interview or provide informed consent, and individuals who later indicated that they no longer wished to participate in subsequent interview rounds.

\subsection{Ethics}

This study was approved by the authors' institutional ethics review committee. Prior to the interviews, the research team explained the purpose of the study, the interview procedures, the methods of data recording, and how the research results would be used. Participation was entirely voluntary. Participants were informed that they could decline to answer any question during the interview and could withdraw from the study at any time without any negative consequences. Interviews were conducted only after informed consent had been obtained from each participant. All interviews were audio-recorded with participants' permission and later transcribed into text for analysis. To protect participants' privacy, all interview materials were anonymized during data processing. Potentially identifiable information, including names, specific workplaces, residential locations, or other identifying details, was removed or replaced. In all publications and reports, participants are referenced only by numerical identifiers. No personally identifiable information is disclosed in any public materials. Participants received financial compensation for their time. Each participant received 50 RMB for the interviews conducted in 2014 and 2017, and 100 RMB for the interviews conducted in 2020 and 2023.

\subsection{Data Collection}

Data were collected through semi-structured interviews conducted in four stages: 2014, 2017, 2020, and 2023. Depending on participants' availability and geographical location, interviews were conducted either face-to-face or online. Each interview lasted approximately 45--90 minutes. With participants' consent, all interviews were audio-recorded and subsequently transcribed for analysis.

Each interview began with questions about participants' background and everyday life, including their age, occupation, educational background, and typical smartphone usage habits. This stage helped researchers understand participants' technological environments and allowed participants to gradually become comfortable with the interview setting.

The interview then moved to participants' everyday experiences using smartphones. Participants were asked to describe how they used mobile applications in daily life, which platforms they used most frequently, and what roles these applications played in their routines. These questions established contextual background for understanding participants' technology use and provided a foundation for later discussions about personal information and privacy.

Building on this foundation, the interview explored participants' understandings of personal information and privacy. Participants were invited to describe what types of information they considered related to themselves and which kinds of information they perceived as more private. This approach allowed participants to define privacy based on their own experiences rather than responding to predefined conceptual categories.

The interview protocol also included timeline elicitation questions that encouraged participants to reflect on their experiences since they first began using smartphones. Participants were asked to recall how technologies, applications, and their own usage practices had changed over time. This approach helped researchers understand the technological contexts participants encountered at different stages and how these contexts shaped their understanding of personal information and privacy.

In addition, the interview included critical incident probes, which encouraged participants to describe specific events or experiences that drew their attention to issues related to personal information, such as a particular application behavior or technological experience that made them reflect on privacy risks. These questions helped reveal the situational triggers behind changes in participants' privacy awareness.

Finally, the interview explored participants' everyday practices for managing personal information. Participants were asked whether they had taken actions such as adjusting app permissions, limiting certain features, or modifying usage habits to reduce privacy risks. They were also asked about challenges encountered when attempting to manage personal information on their smartphones. These questions provided insight into the practical strategies and constraints users experience when navigating privacy issues in mobile computing environments.

The full interview protocol is provided in Appendix B. The same protocol was used across all four interview rounds.

\subsection{Data Analysis}

This study employed an iterative qualitative analysis approach to systematically analyze the interview materials. The analysis was inspired by thematic analysis while incorporating a longitudinal perspective that emphasizes changes in experiences and interpretations over time. The goal of the analysis was not merely to identify users' attitudes toward privacy, but to understand how participants gradually developed and adjusted their interpretations of data access, behavioral tracking, and AI-driven inference through their long-term experiences with smartphones.

All interviews were audio-recorded with participants' consent and transcribed verbatim, and the research team conducted initial data cleaning and de-identification by removing or replacing potentially identifying information such as names, workplaces, or residential locations. The interview materials were organized by participant identifier and interview year, creating a dataset structured around participant to time point units that allowed both cross-sectional comparisons between participants and longitudinal comparisons within the same participant across years. Three researchers independently read the complete set of transcripts and wrote analytic memos to document emerging themes, then conducted open coding, generating initial codes that remained close to participants' original descriptions.

Through repeated comparison and discussion, related codes were gradually grouped into three primary interaction mechanisms that shape users' evolving understandings of privacy: direct data access, behavioral data tracking and utilization, and AI-driven inference. The research team then conducted longitudinal comparative analysis to examine how the same participants described their experiences across different interview rounds, identifying broader patterns in the evolution of privacy awareness and coping strategies alongside technological developments.

Building on these mechanisms, the research team classified each participant's long term trajectory into three stages: an early stage dominated by functional acceptance of data access, a middle stage in which participants explicitly recognized and could describe ongoing behavioral tracking, and a later stage in which participants judged conventional privacy strategies to be largely ineffective. A participant was judged to have undergone the complete three stage evolution only when the three stages successively became the dominant interpretive framework across the corresponding waves; participants who remained within a single stage, or who showed a transition between only two of the three stages, were classified as having undergone a single stage or two stage evolution. \minoredit{Coding} disagreements were resolved through discussion rather than majority vote, with arbitration by the last author when necessary. Where a participant's retrospective account diverged from what they had said in an earlier wave, the contemporaneous transcript was treated as the primary evidence of that participant's view at the time, and the later account was treated as supplementary evidence of their current interpretation rather than a replacement for the original record. Appendix C provides complete definitions of the stages, comprehensive classification criteria, full procedures for resolving coding disagreements and retrospective inconsistencies, and the roles of the researchers.

\section{Findings}

Among the 31 participants, 25 completed all four rounds of interviews. Among these 25, 20 participants experienced the complete three stage evolution: initially, in the early period, accepting default access to private data; then, in the middle period, becoming aware of privacy risks and taking limited management measures; and finally, in the later period, knowing about privacy issues yet gradually feeling powerless and returning to inaction, accounting for 80 percent. Among them, the changes of P3, P5, P7, P12, P15, P16, P19, P23, and P26 were particularly consistent with this trajectory. Taking P3 as an example, their statements show a shift from ``accessing data is reasonable, there is no security issue'' in 2014, to ``there is a problem'' in 2017 when they began to reflect on what had happened, to ``actively turning off certain behaviors'' in 2020, and finally to ``basically doing nothing at all'' in 2023, reflecting the complete evolution through the three stages. Additional cross stage comparisons of other participants are provided in \minoredit{Appendix~\ref{app:additional_quotes}} as further evidence.

P3 (2014): ``I think it is reasonable for the phone to access this kind of data, such as photos and audio. Otherwise how would I interact with it? I do not think this would cause any information security problem.''

P3 (2017): ``I feel that some aspects are problematic. For example, sometimes I see something, and not long after it appears in my recommended feed, so I wonder where they stole this data from. I try to find ways to prevent them from knowing this data.''

P3 (2020): ``Based on my experience, I actively turn off certain behaviors, such as some important permissions.''

P3 (2023): ``Actually, regarding AI, I have also done a lot, but now I basically do nothing, because it cannot be solved at all. However you handle your data, it will still be collected.''

The three stages do not have clear boundaries between them; rather, they represent an overall trend of change. The specific timing at which different participants entered each stage, as well as their degree of engagement, also varied. In addition, P1, P18, and P28 broadly went through the first two stages, but in the 2023 interviews they still reported maintaining necessary privacy protection measures in response to increasingly complex privacy issues, a direction that differed from most other participants. Two further participants also stand out. P29 consistently showed little interest in privacy issues across all four interview waves, while P6 believed from the very start of the study that privacy issues were important yet difficult to resolve, and both of these patterns differ entirely from the evolutionary pattern seen among most users. The specific changes in the three stages are elaborated below.

\subsection{Stage 1: Privacy issues obscured by functional needs}

At the start of the study (2014), most users generally lacked proactive privacy management awareness, which was shown in two aspects: users often treated their smartphones merely as convenient tools for communication, entertainment, and information access in daily life, rarely mentioning privacy issues that might arise during use; and they lacked understanding of privacy issues themselves. The statement from P3 below reflects the view that it was reasonable for a smartphone to obtain data because doing so was necessary for functionality.

\textbf{P3 (2014):} \textit{``I think it's reasonable for my phone to access these kinds of data, like pictures and audio. Otherwise, how am I supposed to interact with it? I don't think this would cause any information security problem.''}

The interviews suggest that this lack of privacy awareness among users was closely tied to direct data access mechanisms. Direct data access was the form of data processing most commonly encountered by users during this period, appearing through permission pop ups and function authorization scenarios that requested access to contacts, photo albums, cameras, microphones, location, and SMS verification codes in order to complete certain functions. Under such requests, most users reported that they focused mainly on completing registration smoothly, quickly entering the application, and using functions without interruption, while very few users mentioned considering the privacy issues involved or adopting any strategies in response.

Taking access to contacts as an example, at this stage most users understood such requests as part of a social application's effort to find acquaintances, import contacts, or facilitate social connections, and accordingly accepted the access. For instance, P15 below understood contact access as a necessary condition for a social app to recommend acquaintances and establish connections.

\textbf{P15 (2014):} \textit{``The social app said it needed access to my contacts so it could recommend people I might know. I thought that was good. If it doesn't know my contacts, how could it recommend anyone?''}

Similarly, taking photo album access as an example, most users regarded this permission as a natural part of sending pictures, uploading profile photos, or posting content, and their typical response was ``I need this to complete the function.'' For instance, P4 below considered album permission a necessary part of the function of sending photos.

\textbf{P4 (2014):} \textit{``When I use WeChat, it asks whether I allow access to my photo album. But if I want to send someone a photo, how could I do that if I don't allow it?''}

Similar situations occurred frequently with other types of data such as camera, microphone, location, and SMS verification codes. For example, P7 below regarded location permission as a precondition for a map service to function properly, while P21 considered microphone access naturally necessary for sending voice messages and making video calls. Although different data types corresponded to different functions, in users' understanding they were generally absorbed into the same operational logic of ``providing the corresponding data in order to complete this function.''

\textbf{P7 (2014):} \textit{``A lot of software asks for location permission when you register. I think that's very reasonable---otherwise what's the point of using maps?''}

\textbf{P21 (2014):} \textit{``If I want to send a voice message or make a video call, then of course I need to give it microphone access. So I just turned it on without really thinking about it.''}

It is worth noting that even when a small number of users began to sense that access to certain types of data might carry unclear risks, this awareness rarely translated into actual privacy management behavior during the early period of the study. Although P28 below mentioned concerns about SMS verification codes and P20 mentioned concerns about permissions such as facial verification for the album, most users ultimately still placed completion of core functions above such concerns, taking almost no corresponding action.

\textbf{P28 (2014):} \textit{``I didn't want to provide the verification code to that banking app, but if I didn't, I couldn't withdraw money. It was just part of the task. Even if there was some privacy risk, what else could I do?''}

\textbf{P20 (2014):} \textit{``When an app asked me for facial verification, I was a little worried. But I really needed to finish registration quickly. And my friends were all registering this way too, so I thought it was probably fine.''}

\secondedit{P20 and P28, both of whom had undergraduate education at baseline, expressed privacy concerns earlier than many participants (as P28 put it, ``maybe I think about it a bit more than others do''). Their 2014 quotations nevertheless show that functional completion still took priority and do not establish that either participant had already adopted proactive management. Education may have shaped the timing of concern, but the interview evidence does not support a causal conclusion.}

\secondedit{In addition, seven participants formed exceptions to the pattern described in this stage. First, four users (P9, P11, P22, P24) differed from most users in this stage, who ``were unaware of the privacy issues involved in smartphone use, or vaguely sensed them without taking action, yet still held some basic understanding of privacy and did not consider privacy issues entirely dismissible.'' The interviews of these four users instead showed a clear lack of even basic understanding of privacy issues, and difficulty recognizing their importance at all (for example, P11 said, ``I am just an ordinary person, what privacy is there to speak of,'' and P24 said, ``I have not done anything wrong, why should I be afraid of privacy leaks''), which tended to delay their entry into the next stage relative to the overall group. Second, two participants with relatively high education (P2 and P8, both with undergraduate backgrounds) had already reached the next stage by the time of the 2014 interviews, ahead of the overall group, and had already taken partial measures against smartphone privacy issues at this stage (for example, P2 mentioned, ``I actively review and turn off permissions I consider unnecessary''). Third, P6's pattern differed entirely from all other participants. P6 had a relatively high level of education and technical literacy along with an understanding of privacy issues, yet adopted an attitude of resignation that most other participants did not display until later interview waves: ``I admit privacy issues are important, and certain phone functions do leak privacy, but speaking from my professional understanding, this is a problem that cannot be solved. There is too much information out there. Just give up.''}

\subsection{Stage 2: The emergence of privacy concerns and limited management}

As time passed, more and more participants began to realize that personal information might be continuously recorded and used during smartphone use, a shift that mainly occurred between the end of the 2014 interviews and the 2020 interviews. They increasingly used expressions such as ``worried,'' ``not entirely at ease,'' or ``I think about this more now.'' Privacy gradually shifted from an abstract concept into a real issue requiring attention. For example, both P5 and P13 below expressed dissatisfaction about the leakage of their own privacy.

\textbf{P5 (2017):} \textit{``I've realized how much information my phone is stealing, and what this information can be used for---for example, inferring my preferences from my browsing. Now I try to resist this theft of information.''}

\textbf{P13 (2020):} \textit{``The information I posted on social media last time was misused. I don't know where they got my phone number, but ever since I posted something about fishing, I've been getting spam calls every day trying to sell me fishing rods. It made me really angry.''}

The interviews suggest that this shift was influenced by background data processing along with feedback and visibility, as well as by social consequences. Unlike direct data access mechanisms, behavioral data tracking is embedded in participants' long term smartphone use. Mobile platforms continuously record users' searches, browsing, dwell time, clicks, swipes, location movement, and social interactions in the background, and then convert this data into feedback delivered to users, including behavioral feedback and system visibility.

The first is behavioral feedback, which mostly reflects prior behavioral data back to users through advertisements, recommendation feeds, and related content. It is precisely through this correspondence between ``past behavior'' and ``current content'' that users gradually came to suspect, reflect on, and infer that continuous data recording and use was taking place in the background.

\minoredit{\textbf{P15 (2020):} \textit{``One minute I'm looking at fishing stuff in App A, and one minute later App B is recommending fishing rods. This is definitely a problem. They must be collecting my data and using it, and I had no idea at all.''}}

\textbf{P9 (2017):} \textit{``Sometimes it recommends people to me and says I might know them, but I never told it about those relationships.''}

\minoredit{In the advertising example P15 mentioned above, when fishing related behavior repeatedly corresponded to subsequent content, users began to suspect that this behavior was being continuously retained and used by the platform. P9, meanwhile, raised questions about how the platform obtained relational information, and began to infer that the platform might be integrating contacts, mutual connections, or other relational clues. Together, these cases illustrate how behavioral feedback prompted suspicion about data collection and use, contributing to the shift in privacy awareness and strategy. An additional recommendation example from P14 is provided in Appendix~\ref{app:additional_quotes}.}

Second, system visibility directly brought previously hidden data access processes into view for users. Through interface prompts, access logs, or system status indicators, users were able to observe the specific timing, targets, and frequency of background data calls, making what had once been an abstract process of data collection into something concrete and perceptible, which prompted users to reconsider the data access they had previously accepted by default.

The most direct change came from device indicators. When an app called the microphone or camera, the system would often display a corresponding indicator in real time, making participants aware that data access was not merely something that happened at the permission authorization stage, but something that actually occurred at a specific moment. For example, P12 below described the immediate alertness triggered by visible real time data access.

\textbf{P12 (2017):} \textit{``When that indicator lights up and tells me something is being recorded, I always feel a bit tense and uncomfortable.''}

Beyond real time prompts, access logs further changed how participants understood the persistence of background data processing. For example, P17 below mentioned that permission logs showed the background might still repeatedly call certain permissions even without active operation, which prompted P17 to reconsider background data processing and privacy issues.

\textbf{P17 (2020):} \textit{``I thought that if I wasn't using it, then it wouldn't touch those things [microphone, photo library]. But then I looked at the records and found that it had accessed them many times.''}

\minoredit{A further example of interface reminders reinforcing this awareness (P16) is provided in Appendix~\ref{app:additional_quotes}.}

Finally, a small number of users mentioned relatively serious consequences of privacy leakage that became a direct cause of changes in their thinking and behavior, mainly involving fraud and harassment. For example, P27 and P18 below illustrate these consequences from different angles.

P27 (2020): ``Once, because that information was leaked, a scammer used my information to defraud a large amount of money from my child. That experience made me reflect deeply.''

P18 (2020): ``I left my phone number, ID number, and bank card information on that website, and later my ID was somehow used to take out a mortgage loan without my knowledge. The police even came looking for me over it.''

As users' attention to privacy issues continued to grow during this stage, most participants began to proactively adjust their usage habits. However, this change did not take the form of leaving platforms or completely cutting off data flow, but rather appeared as limited adjustments to certain behaviors and data boundaries, mainly in the following forms.

\minoredit{The first is reducing visible behavior. P19 and P22, quoted below, described limiting searches and likes to reduce the behavioral signals available to platforms. A related example of reduced public expression (P14) is provided in Appendix~\ref{app:additional_quotes}.}

\textbf{P19 (2017):} \textit{``Before, I searched whatever I wanted. Now there are some things I think twice about, because once you search them, it keeps recommending them afterward. Some kinds of information are risky simply because the algorithm knows about them.''}

\textbf{P22 (2020):} \textit{``Now sometimes when I see history videos that interest me, I don't even like them anymore, otherwise it keeps pushing me further in that direction. My major isn't history, and I worry that if my whole phone is full of history content, it will affect my studying.''}

In total, 16 participants across the 2017 and 2020 interviews described reducing public expression on platforms in this way, for example by avoiding sensitive topics in comment sections or sharing fewer details about everyday life, in order to reduce the amount of behavioral information that platforms could directly capture.

Besides reducing behavioral exposure, 13 participants also tried to actively disrupt the platform's existing data judgments, hoping to weaken profiles that had already formed, by browsing different types of content, quickly swiping past certain recommendations, deleting search history, clearing browsing history, or even occasionally uninstalling and reinstalling apps in hopes of breaking the stable recommendation patterns the system had developed. For example, P4 below mentioned uninstalling and reinstalling an app, and P16 mentioned deleting search history.

P4 (2020): ``I keep uninstalling and reinstalling Douyin. I am not sure it really works, but I do feel that the recommended content changes a lot every time I reinstall it, and I quite like that.''

P16 (2017): ``After I use the browser, if the data is fairly sensitive, I delete all of it, since that feels safer.''

Meanwhile, another group of users began to redraw data boundaries across different contexts, rather than continuing to let all data concentrate under the same identity or platform. Through distinguishing platform purposes and using different accounts, they tried to limit the continuous integration of different types of data.

\textbf{P15 (2020):} \textit{``I don't share work-related information on TikTok or Weibo. I use our company's internal system instead. Sometimes I still feel there is some privacy risk, but overall I think it solves a lot of the problem.''}

\textbf{P17 (2020):} \textit{``There are some things I won't look at on my main account, for instance those boring gossip videos and things like that. I'll switch to a smaller account or just browse without logging in.''}

Beyond behavioral adjustments, around two thirds of participants also became more active in using system settings. For example, they disabled unnecessary permissions, changed location access to ``only while using the app,'' or restricted access to contacts, microphones, and photo libraries.

\textbf{P20 (2020):} \textit{``I turned off every permission that could be turned off. If I need it later, I'll think about whether to turn it back on.''}

These participants no longer assumed by default that all behavioral data should continuously flow into the same platform, account, or device, and instead began actively redrawing data boundaries in hopes of limiting the platform's ability to continuously integrate personal information. This shows that participants' concerns had expanded to include limiting data linkage and identity integration.

It is also worth noting that many users mentioned reinterpreting the early data access mechanisms.

\textbf{P10 (2020):} \textit{``Looking back now, I really had no idea how much information I was leaking. Luckily I realized it now---I've turned off every permission I can.''}

However, most users also mentioned the limitations of these strategies and their own frustration. They noted that these strategies could often only affect part of the data collected, without truly interrupting continuous background tracking.

\textbf{P16 (2020):} \textit{``Honestly, of course there are times when I slack off. These privacy strategies add a lot of burden. When I'm tired, things I think I should do---like deleting browsing history---I might just forget.''}

\textbf{P26 (2020):} \textit{``I can control what I browse and whether I click on a video, but later I realized that even the amount of time I stay on a page seems to be used by the algorithm. That's beyond what I can intervene in. I can't control my dwell time and scrolling rhythm too---then I wouldn't even be living.''}

P16 pointed out that these strategies required continuous extra effort and were difficult to sustain over the long term, while P26 mentioned believing that users could not truly and effectively intervene in the collection of private data. Both reflect users' growing awareness of the limits of their strategies, which is also one important reason for the negative emotions and behavior that emerged among most users in the next stage. Even so, at this stage these users still continued to take measures in an effort to avoid privacy problems.

It is also worth noting that the four users mentioned in the previous stage as being slower to recognize privacy issues (P9, P11, P22, P24) all mentioned, at this stage, concerns about smartphone privacy issues and strategies they had adopted in response (for example, P11 said in 2020, ``Only now do I feel there is a problem. It, meaning the phone, knows a lot about me that I do not want it to know, so now I have turned off some permissions''). Meanwhile, P2 and P8 mentioned in the previous stage, as well as P13 who emerged in this stage, already displayed clearly negative emotions about privacy issues (entering the final stage earlier than the overall group, while most users did not begin showing clear negative emotions and behavior until AI emerged near the end of the study). For example, P2 said in 2020, ``There is nothing to be done, that is just how it is. We can never win against technology. These issues, meaning privacy, are destined to require compromise.''

\subsection{Stage 3: The complication of privacy issues and users' sense of helplessness}

By the later period of the study (2020 to 2023), and especially as artificial intelligence became increasingly deeply embedded in mobile ecosystems from late 2022 through late 2023, most users expressed even greater concern about privacy issues and mentioned some new strategies they had adopted. However, they also described the severe limitations of these concerns and strategies, such that most users ultimately came to a state in which they recognized the importance of privacy issues and clearly understood that their phones were collecting private data, yet felt unable to do anything about it, found it difficult to take proactive action, and gradually fell into passive resignation.

\minoredit{First, most users explicitly mentioned concern about how the ongoing development of technology, especially AI, had made privacy issues more complicated. An additional statement of this concern from P12 is provided in Appendix~\ref{app:additional_quotes}.}

This concern was mostly triggered by two capabilities of artificial intelligence. On one hand, natural interaction made communication between people and systems increasingly resemble everyday conversation, leading users to worry that the conversation process itself would continuously reveal more personal information (as seen in P12 below). On the other hand, users realized that the system was not making judgments based on a single input, but was continuously integrating information from multiple sources to form an increasingly complete profile of the individual (as seen in P16 below).

P12 (2023): ``It feels so much like a real person that it is easy to forget you are not talking to a machine. Sometimes as I keep talking, I end up saying too much.''

P16 (2023): ``I increasingly feel that I am not talking to an AI on my phone, but to a person on the other side of the screen who can infer things about me from the smallest details. That feels really frightening.''

The capacity of AI systems to draw inferences from multiple data streams is a general property of the underlying technology. The extent to which these streams can be integrated into a single identifiable profile, however, depends heavily on whether such data are held within the same platform ecosystem, a condition closely tied to the consolidation of communication, payment, and service functions characteristic of super app architectures~\cite{jia2022superapps}.

Faced with this new privacy risk, most users did not immediately give up, but instead tried to further adjust their usage habits. First, some users began proactively limiting the information they input, including reducing the forms of input, reducing the amount of information shared, and avoiding private information. For example, P7 below mentioned reducing the use of voice input, and P3 mentioned avoiding providing complete background information across multiple turns of conversation, hoping to limit the contextual information the system could continuously accumulate.

\textbf{P7 (2023):} \textit{``If I'm just checking the weather, I'll use voice. But for some questions, I still type them. I'm afraid AI will know too much.''}

\textbf{P3 (2023):} \textit{``I won't keep asking a lot of things in the same window, because it feels like it will remember more and more.''}

However, more and more users began to realize that these localized adjustments could only affect a very limited scope and could hardly change the overall situation. P7 soon added, ``I stopped doing that pretty quickly, because I felt that compared with all the information AI already collects about me, one less voice message really makes no difference.'' P3 added, ``I feel that not describing things in the same window is basically self deception. Of course I have no clear evidence, but based on how it feels, I think AI still knows everything, and I have not seen any real benefit from doing this.''

P11's description below further supports the widely shared sense among users that such actions were ineffective. P11 believed that as long as one used digital devices, data leakage was unavoidable.

\textbf{P11 (2023):} \textit{``Surveillance in the digital age is ubiquitous; merely making localized adjustments to your information won't be effective. You simply cannot completely isolate yourself from data---I mean, you know how it is: for instance, even if you choose not to use a specific app for a particular task, you can't realistically go completely without using apps altogether, right? Consequently, there will always be information channels through which your data can leak out.''}

P11's sense of being unable to opt out stems in large part from the high degree of bundling among payment, social, mobility, and identity services within a small number of super apps in China~\cite{jia2022superapps, plantin2019wechat}. In contexts where these services are distributed across multiple independent and substitutable platforms, the exit costs users face, and the resulting sense of powerlessness, may differ in intensity.

Beyond users' difficulty in seeing any benefit from their actions, as mentioned above, two other factors accelerated users' abandonment of action. The first is technical complexity. As technology advanced, most users found it increasingly difficult to judge exactly what data the system was recording and what conclusions it was drawing from that data. This high degree of uncertainty steadily eroded their confidence in managing their own privacy. For example, P23 below mentioned not knowing how much data the phone actually recorded.

\textbf{P23 (2023):} \textit{``You can say a little less, but how much it has actually recorded---you don't really know.''}

The second is the accumulation of burden. The emergence of AI driven inference did not replace earlier data extraction mechanisms, but was layered on top of them. Near the end of the study, users had to deal with direct permission requests, continuous behavioral tracking, and opaque algorithmic inference all at once. This layering effect created an unprecedented cognitive burden.

\textbf{P11 (2023):} \textit{``Now, not only do I have to turn off my phone's location services, but I also have to guard against accidentally tapping on recommendations; even speaking to my voice assistant requires extreme caution. There are simply too many things to keep track of---it's exhausting, and it feels utterly impossible to stay fully protected.''}

This sense of cognitive overload from managing multiple tracking mechanisms simultaneously may be common wherever users face permission systems, behavioral tracking, and AI assistants together, though the intensity of this burden may vary with how concentrated these functions are within a single platform~\cite{jia2022superapps}.

\minoredit{It is worth noting that a small number of users (such as P4, P10, and P17) still mentioned certain relatively basic privacy protection measures, which were related to the understanding of smartphone privacy they had accumulated during the first two stages. Overall, however, these behaviors were difficult to characterize as active or effective resistance. P10, quoted below, illustrates these limited measures; a related account from P17 is provided in Appendix~\ref{app:additional_quotes}.}

P10 (2023): ``Honestly there isn't much left to do. Overall, I feel like I've basically stopped worrying about privacy issues, but I think compared to 2016 or 2017, I have definitely made some progress, I mean in terms of strategy. There are some particularly important things I won't upload to certain places anymore, but that's about it. For most other things, there's nothing I can really do.''

By contrast, a small number of users did not participate in this stage of evolution (P1, P18, and P28). All three mentioned that they still placed importance on smartphone privacy issues and actively adopted corresponding strategies. Among them, P1 and P28 were driven by their own knowledge, while P18 mentioned that the consequences of a privacy breach had increased their level of concern about privacy issues. For example, P1, quoted below, mentioned believing that learning could enhance one's ability to guard against potential privacy breaches, while P18 mentioned the impact of a personal experience of privacy breach.

P1 (2023): ``Even though information is getting more and more complicated, I still think there are ways to avoid it, I mean privacy breaches. For instance, I can go and learn how AI actually works. Either way, if it wants to collect my information, it still has to rely on sensors, right? So if I don't carry my phone with me for a certain period of time, wouldn't it be much harder for my information to leak during that time? I think this is really a matter of attitude, I mean, we need to be willing to keep learning.''

P18 (2023): ``And you know, I went through a serious privacy breach before, so now I'm especially careful about this. I worry about the past happening again, honestly, it's hard for people to really understand something like this unless it happens to them.''

It is worth noting that the statements of P1, P18, and P28 also all acknowledged that the privacy awareness and measures maintained at this stage placed a considerable burden on them, which is similar to the pressure mentioned by other participants above.

P28 (2023): ``I often feel like I'm just one step away from breaking down. Every day, in order to protect my information, there are all kinds of messy things I have to do, like repeatedly turning permissions off and on. There are a lot of functions I can't use normally. Honestly, sometimes I feel like I'm about to collapse from exhaustion, but I can't give up. I sometimes wonder if I'm being too paranoid.''

\subsection{Attrition and Subgroup Divergence}

\paragraph{4.4.1 Attrition.}

Of the 31 participants, 25 completed all four rounds of interviews, and 6 withdrew partway through. P31 withdrew for personal health reasons, completing only the first round (2014). P24 withdrew due to work, completing only the first two rounds (2014, 2017). P30 relocated for work and formally invoked the study protocol's ``withdrawal without reason'' clause, completing only the first two rounds (2014, 2017). Both P20 and P21 withdrew due to increased work demands and persistent scheduling constraints, each completing the first three rounds (2014, 2017, 2020). P27 was absent in 2017 due to personal health issues, resumed participation in 2020, and was absent again in 2023, completing only the first and third rounds. After the study concluded, the research team attempted to recontact these six withdrawn participants to obtain their most recent privacy attitudes. P20, P21, P24, and P31 were all successfully reached, and their attitudes toward privacy largely continued the overall evolutionary trend seen among the retained participants (from early default acceptance, to limited adjustment in the middle period, to high vigilance combined with a sense of helplessness in the later period), consistent with the overall three stage trajectory reported in this paper. For example, P20 said, ``A few years ago I tried hard to avoid privacy leakage, but looking at it now, it really did not seem to have much effect, so I am not very motivated to act anymore.'' This suggests that their withdrawal was mainly due to time and work scheduling constraints rather than a loss of interest in privacy issues. P27 and P30 could not be recontacted, so their subsequent attitudes remain unknown based on the available data.

\paragraph{4.4.2 Divergence.}

\begin{figure}[htbp]
    \centering
    \includegraphics[width=0.8\textwidth]{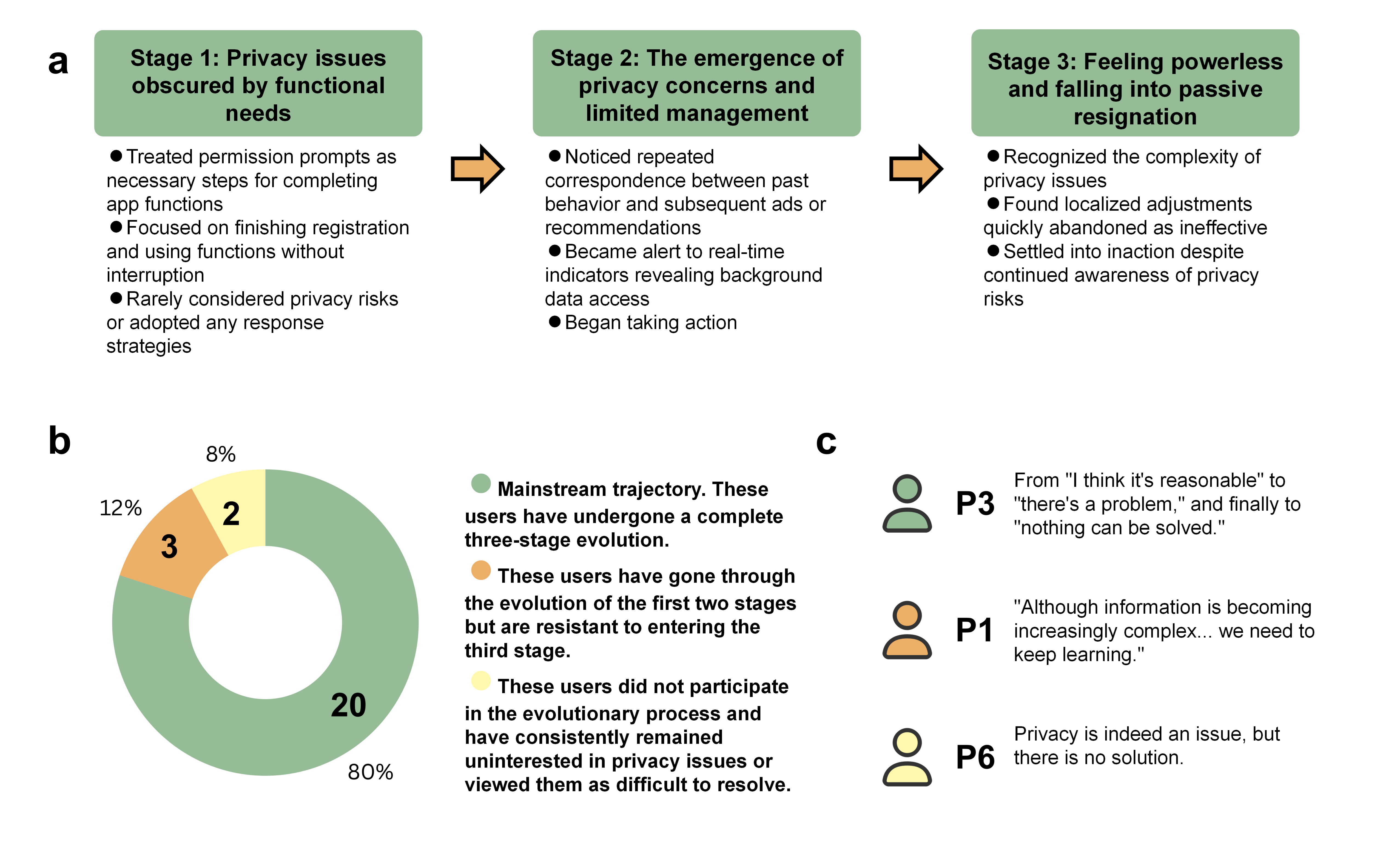}
    \caption{a: Schematic of the three stages; b: Differentiation results for the 25 participants who completed all four interviews; c: Representative examples of participants.}
    \label{fig:figure1}
\end{figure}

Two users showed evolutionary patterns entirely different from the overall three stage pattern described above. First, P29 consistently showed no interest in privacy issues throughout (``Privacy is not a particularly important issue,'' 2014; ``Am I really the only one who does not care about privacy,'' 2017; ``I have not taken any action, I have always been like this, and I do not see anything wrong with it,'' 2020; and ``I think my view has not changed since the earlier interviews.''). Second, P6 consistently believed privacy issues could not be avoided, repeatedly using expressions such as ``a problem that cannot be beaten'' and ``it truly is a problem, but it cannot be solved'' between 2014 and 2023. However, P6's views showed some internal tension. In the final interview in 2023, P6 briefly mentioned ``doing what little one can,'' acknowledging that some privacy protective behaviors did have value, and referred to ``limited protective actions.'' Overall, P6 did not go through the kind of change most other users experienced, or at least changed at a far slower pace than all other participants.

In addition, three participants broadly followed the pattern observed among most users, but diverged in the final stage. These included P1, P18, and P28, who mentioned actively engaging in certain proactive privacy protection behaviors only during the final stage. Although these behaviors placed a considerable burden on them, this stands in contrast to other participants. All three of these participants also possessed relatively high levels of education and digital literacy, and among them, P18 in particular mentioned having previously experienced a serious privacy breach that led to identity theft, which was an important reason behind P18's behavior during the final stage.

As shown in Figure~\ref{fig:figure1}, \minoredit{the} remaining 20 participants all completed the overall three-stage evolution, albeit at different speeds. For instance, P2 and P8 evolved noticeably faster than the overall group, while P9, P11, and P22 evolved noticeably more slowly than the overall group. In addition, P4, P10, and P17 mentioned during the final stage that they still maintained ``very limited privacy protection behaviors (P4)'', which lends further support to the diversity within this group; that is, users did not entirely abandon all strategies in the final stage, but instead exhibited an uneven pattern in which actions had contracted significantly overall, within a broadly passive state.

\paragraph{4.4.3 Subgroup analysis.}

Subgroup comparisons were conducted among the 25 participants who completed all four rounds of interviews to examine similarities and differences in long term privacy attitude evolution across demographic backgrounds. The comparisons showed that gender, urban or rural background, education level, and age had no clear effect on whether a participant ultimately completed the full three stage evolution. By gender, 11 of 13 men (84.6\%) and 9 of 12 women (75\%) experienced the complete evolution. By urban or rural background, 12 of 14 rural participants (85.7\%) and 8 of 11 urban participants (72.7\%) experienced the complete evolution. Grouped by education level in 2014 (literate, primary school, or middle school as the lower education group, 6 people; high school or above as the medium to high education group, 19 people), 5 of 6 (83.3\%) and 15 of 19 (78.9\%) respectively experienced the complete evolution. Although some numerical differences existed across groups, in every group the complete three stage trajectory remained the dominant pattern, and each group included cases of earlier entry into the next stage, delayed formation of privacy awareness, and individual long term deviation from the trajectory, with no stable or consistent grouping pattern observed. The same held true for age, with participants who entered the next stage relatively quickly and participants who deviated from the overall trajectory over the long term appearing across all age groups, and no clear correspondence between age and evolutionary pattern was observed.

However, the two education groups may have differed somewhat in the timing of entry into each stage. In the medium to high education group, P2 and P8 had already begun actively reviewing and turning off certain permissions by 2014, entering the second stage earlier than most participants, and P20 and P28 also showed concern about certain data access at the early stage of the study. By contrast, P11, P22, and P24 in the lower education group still lacked sufficient awareness of privacy risks at the start of the study, and only gradually developed clearer privacy awareness and began adopting corresponding measures in later interviews. In the second stage as well, users with higher education tended to mention beginning their privacy strategies earlier. However, given the small sample size, no confident conclusion can be drawn on this point.

Compared with demographic characteristics, individual differences at the level of trajectory were more pronounced. Among the 25 participants who completed all interviews, 20 experienced the complete three stage evolution. Three participants generally went through the first two stages but continued to take some privacy protective measures in 2023, without fully abandoning proactive management. P29 consistently lacked interest in privacy issues throughout. P6, from the start of the study, consistently believed that privacy issues were objectively real but almost impossible to solve, and did not overall experience the staged changes shown by most participants. This suggests that the three stage model can capture the long term evolution experienced by most participants, but individuals differ mainly in the timing of their entry into each stage and in whether they ultimately continued to persist in privacy management.

An exploratory qualitative comparison suggests that platform dependency may shape individual trajectories, though its influence appears complex. We identify two contrasting patterns here. First, P29 (sex worker, later cleaner), whose livelihood was closely tied to platform based client acquisition and transactions, showed no engagement with privacy concern across all four waves. P29 instead emphasized needing certain functions and having no capacity to worry about privacy, with many of these functions tied to occupational needs such as greater social visibility. P29 stated, ``I need a certain amount of exposure for my work. I do not really have room to worry about privacy on top of that.'' P13's account showed a related pattern, describing, ``Honestly, sometimes I really want to deal with these privacy issues, especially the misuse problem. But I need social media to build my visibility. It is a bit contradictory.'' In both cases, platform dependency appears to have muted concern about privacy to some degree. Second, P11 (farmer), whose occupation involved comparatively little reliance on platform based transactions, entered the vigilance stage markedly later than the overall cohort. In 2020, P11 remarked, ``I have not used it much, so it is only now that I am starting to think there might be some kind of problem.'' Lower platform dependency corresponded to more limited usage experience, which in turn appears to have delayed P11's recognition of privacy risk.

\paragraph{4.4.4 The Role of China's Platform Structure in Shaping the Findings.}
Some of the findings reported in Sections 4.2 through 4.3 were shaped by structural conditions specific to China's mobile platform environment, namely platform concentration and the bundling of services under a super app architecture. As shown in Appendix~E, AI driven inference across multiple data sources was able to produce the sense of a complete profile that participants described precisely because voice, chat, location, and device data already converged within the same account ecosystem, a structural consequence of the super app architecture. Participants' sense of being unable to opt out was similarly shaped by this structural condition: because payment, social, mobility, and identity verification services were bundled within a small number of apps, users found it more difficult to limit data flows by switching platforms, which constrained their exit options. This suggests that the pressure and sense of powerlessness described in Stage 3 above may have been amplified, to some degree, by this structural feature of China's mobile platform environment, and that the same intensity of experience might not arise in regions with lower platform concentration or weaker service bundling. By contrast, the functional acceptance of permission requests, the vigilance triggered by behavioral feedback, the reassessment prompted by system visibility cues, and the limited adjustment coping strategies describe how users came to notice and respond to data practices as a cognitive process in its own right; their emergence does not presuppose that services are concentrated on a single platform, and we find no clear indication that they were shaped by China's platform structural conditions. It should be emphasized that this classification reflects an analytic judgment rather than a conclusion drawn from cross context empirical comparison.

\minoredit{Taken together, the findings trace a dominant shift from treating permissions as functional requirements, through vigilance and localized adjustments prompted by tracking feedback, to resignation as accumulated data and AI-mediated inference appeared to exceed individual control. This pattern characterized 20 of the 25 participants who completed all four waves, but it was neither uniform nor inevitable: P1, P18, and P28 sustained active management, P29 remained largely unconcerned, and P6 expressed powerlessness from the outset. Differences in timing and continued protective practices show that resignation did not simply mean indifference or the abandonment of every precaution. Across these trajectories, the central tension was between expanding awareness of data use and participants' perceived ability to act, connecting the findings to the discussion of cumulative infrastructure exposure.}

\section{Discussion}

This discussion argues one claim: mobile privacy resignation emerged through cumulative infrastructure exposure, not apathy. Participants first interpreted permissions as functional access to local device resources, later reinterpreted them through behavioral tracking and platform feedback, and finally experienced AI inference as ecosystem-level profiling exceeding individual agency. The contribution thus goes beyond describing changing privacy attitudes in China: it explains how ubiquitous computing infrastructures make privacy risk progressively legible while making effective action harder, linking theory, design, and scope through accumulated sensing, tracking, synchronization, cross-device flows, platform dependency, and AI inference.

\subsection{Infrastructure Exposure and Longitudinal Privacy Theory}

\secondedit{We define cumulative infrastructure exposure as a process involving three linked operations: data-capable layers accumulate; later feedback reveals previously hidden linkages; and users retrospectively revalue earlier actions as inputs to broader profiling. Its analytical entry point precedes related constructs by beginning with functional acceptance before privacy risks become perceptible. Its core drivers are layer accumulation, cognitive lag, retrospective revaluation, and an agency mismatch between ecosystem-scale risks and local controls. Its temporal claim is that later tracking and inference layers change the meaning of earlier actions, explaining why seemingly functional permission grants are later reclassified as profiling inputs.}

\secondedit{By contrast, privacy fatigue explains why accumulated decision burdens reduce willingness to manage privacy, even within a relatively stable system \citep{choi2018role}. Privacy cynicism explains why institutional mistrust and powerlessness coexist with continued use, while surveillance realism explains why discomfort becomes normalized when alternatives appear constrained \citep{hoffmann2016privacy, dencik2017advent}. Digital resignation identifies the outcome in which users still desire control but feel unable to act under asymmetric data power \citep{draper2019corporate}. These concepts thus primarily characterize attitudes, states, or outcomes after risks, mistrust, constrained alternatives, or powerlessness have become salient. Cumulative infrastructure exposure is not another label for low agency; it specifies an upstream, temporally ordered mechanism that can contribute to fatigue, cynicism, surveillance realism, and, in this study, digital resignation. It predicts resignation only when revealed infrastructural capacity grows faster than actionable control; where provenance remains attributable and controls scale with it, exposure may instead produce sustained adaptation.}

Divergent trajectories provide boundary checks on this mechanism. P6 recognized infrastructural accumulation early and reached resignation before most participants, suggesting that technical understanding can shorten cognitive lag without restoring agency. P1, P18, and P28 sustained adaptation when learning or breach experience kept some practices attributable and actionable. By contrast, P29's livelihood dependence on platform visibility muted vigilance. These cases support the conditional claim that exposure yields resignation when perceived infrastructural capacity outpaces actionable control.

The longitudinal perspective does not show that classic privacy theories are wrong. It shows that their central constructs become unstable when privacy risks are revealed only after infrastructures accumulate. Privacy calculus is complicated because benefits are immediate, while costs may become visible only later, after tracking, synchronization, and inference layers connect earlier traces \citep{dinev2006extended}. The privacy paradox need not indicate inconsistency or irrationality; the attitude-behavior gap can emerge because risk visibility, dependency, and exit costs change over time \citep{barnes2006privacy}. Communication privacy management and boundary regulation also extend beyond interpersonal disclosure: participants tried to preserve boundaries across apps, accounts, devices, platforms, recommendation systems, synchronization, and AI memory \citep{petronio2002boundaries}. For UbiComp privacy theory, the unit of analysis shifts from an isolated choice to the temporal relation between user action and infrastructural accumulation.

This retrospective revision is clearest in relation to contextual integrity. Nissenbaum's account locates privacy violations in inappropriate information flows relative to contextual norms \citep{nissenbaum2004privacy}. Our findings suggest that ubiquitous inference can undermine users' ability to identify the relevant flow in the first place. Participants could often sense that information from one setting had resurfaced in another, but they could not determine whether the input came from a permission, search, dwell time, advertising identifier, synchronized account, connected device, or AI conversation. The problem was not only that information moved against contextual expectations. It was that the context of origin became difficult to recognize because inference assembled meaning across multiple contexts.

The theoretical implication is that contextual integrity depends on practical attribution. Users continued to value separation among work, family, health, leisure, and social expression, as reflected in their attempts to use secondary accounts, guest modes, separate devices, and curtailed expression. Resignation did not signal abandonment of contextual norms; it signaled the erosion of users' capacity to perceive, attribute, and contest violations of those norms. Infrastructure exposure thus names the mechanism by which ubiquitous inference can weaken contextual judgment before it produces explicit withdrawal. For IMWUT, this positions mobile privacy resignation as an infrastructural outcome of evolving sensing and inference systems rather than a stable psychological disposition.

\subsection{From Access Control to Longitudinal Privacy Support}

The design implication is that UbiComp privacy research should be more skeptical of permission-centrism. Work on notices, nudges, labels, dashboards, and runtime indicators has improved how users learn about data access \citep{schaub2015design, zhang2016privacy, almuhimedi2015your, kelley2010standardizing, mhaidli2020listen}, and such work remains valuable because permission interfaces are often poorly understood \citep{felt2012android} and app behavior can exceed expectations \citep{shklovski2014leakiness}. Yet the trajectory in this study shows that one-time decisions about raw resource access are inadequate for risks produced across the longer lifecycle of access, tracking, synchronization, personalization, and inference. Better permission UX targets an important but early moment in that lifecycle.

Transparency also needs to be paired with agency at the same infrastructural level as the problem. Runtime indicators and access histories made background activity visible, but visibility alone did not give participants ways to respond when tracking and inference appeared to operate across accounts, platforms, devices, and AI services. In such cases, transparency can inform users while intensifying the sense that only local controls are available for ecosystem-scale data flows. The design challenge is therefore not simply to expose more events, but to connect visibility with actionable controls over access, inference, synchronization, and personalization.

\secondedit{First, inferential visibility could be implemented as an OS/platform inference-provenance ledger. Apps and model services would attach machine-readable declarations to an inference request---its purpose, predicted attribute, source namespaces, recipient, and retention---while the OS adds observable sensor and account accesses. A runtime card could then state, for example, ``This response uses the current prompt and Health memory to infer a medication schedule; retained for 30 days,'' with controls to disable a source and rerun. Prior transparency systems provide foundations for tracing personal-data processing \citep{li2017privacystreams, barbosa2021who}, while work on sensitive AI inferences motivates exposing derived attributes, not only raw access \citep{asthana2024know}. Because an OS cannot fully observe opaque server-side inference, undeclared remote processing should be labeled ``unknown,'' not presented as absent.}

\secondedit{Second, context isolation could use OS-level context profiles backed by separate identifiers, storage keys, synchronization scopes, and model-memory namespaces. Each data event inherits the active context; APIs deny cross-context reads by default. A purpose-bound, expiring bridge would permit a specific transfer---for example, sharing one Work calendar slot with a Health task without exposing work email history, the advertising identifier, or persistent Work memory. This operationalizes contextual integrity as enforceable information-flow rules \citep{nissenbaum2004privacy, barth2006privacy} and turns participants' secondary accounts, guest modes, and separate devices into infrastructural support rather than fragile workarounds.}

\secondedit{Third, a local policy agent could compile user rules into controls over the ledger and context bridges, extending personalized permission assistants beyond one-off access decisions \citep{liu2016follow}. Consider a user who opens a Health context and asks an AI assistant to draft a note to a doctor. Before execution, the interface lists the prompt, selected medical file, and Health memory as active, while Work email, location, advertising profile, and leisure chats remain off. If the assistant requests a calendar or contact, the agent previews the requested bridge, purpose, expiry, and a degraded alternative; the user can approve once or continue without it. Afterward, the token expires, the action appears in an audit log, and retained memory can be reviewed or deleted. This walkthrough couples visibility, isolation, and delegated action at the point where a context boundary would otherwise be crossed.}

\secondedit{These components are especially relevant to LLM-based agents, although this is a forward-looking extension of the 2014--2023 findings. Persistent memory, multimodal input, and tool use can join contexts and expose private interactions \citep{wang2025agentmemory, chen2025multimodalprivacy, debenedetti2024agentdojo}. The ledger should therefore record data and tool provenance; context profiles should scope memory and tool tokens; and the policy agent should run locally where possible, retain minimal preference data, preview consequential or irreversible actions, and support override and recovery. Service attestations and independent audits remain necessary for opaque server pipelines, and the interface must expose uncertainty rather than imply complete enforcement.}

Persistent agent memory sharpens the mechanism because it can retroactively connect disclosures across sessions and contexts, compressing retrospective revaluation into the routine operation of one assistant. Tool use adds a distinct risk: an agent may not only infer across contexts but also act across applications. Privacy support must therefore govern what an agent may do, with consequential actions previewed, bounded, and reversible \citep{debenedetti2024agentdojo}.

These proposals also have limits. A provenance ledger depends on service declarations, cannot fully observe server-side inference, and creates a sensitive record of its own; frequent runtime cards could reproduce privacy fatigue. Context isolation sacrifices some integration and conflicts with platform incentives, while a policy agent requires trust and may weaken users' situated awareness. Ledgers should therefore be strongly protected, routine events aggregated, cross-context access narrowly scoped, and consequential actions made reviewable and reversible. Adoption will ultimately depend on governance and business incentives as well as technical feasibility.

\subsection{Scope, Transferability, and Future Work}

China is best understood here as a dense case of mobile infrastructure exposure, not as a setting from which universal claims should be directly generalized. Several scope conditions matter. Mobile life in China is deeply organized through platform ecosystems that combine communication, payment, mobility, entertainment, work, identity verification, and public services, making platform dependency especially likely to foreclose exit. State surveillance infrastructures and private platforms are co-constitutive in ways that shape privacy expectations and practical exit. The Great Firewall limits access to alternative platform ecosystems, and the Personal Information Protection Law operates within a cybersecurity governance architecture that differs from GDPR-style rights, redress, and regulatory constraint. These conditions intensify the infrastructural density through which resignation developed, but are not equally necessary for each mechanism.

At the same time, the study identifies mechanisms that can travel under specified structural conditions. Platform dependency does not require China, but does require high platform concentration, bundled essential services, weak substitutes, or institutional constraints on exit; the Chinese case combines all four. Cognitive overload can arise wherever users must manage permissions, behavioral traces, account boundaries, synchronization, and AI inputs across devices. Opaque inference can erode contextual judgment wherever personalization crosses contexts without attributable data flows, although state-platform co-constitution can raise the stakes of opacity. Fatigue can become cynicism when new layers of sensing, tracking, and inference repeatedly make prior strategies feel ineffective; China-specific limits on substitutes, redress, and exit intensify rather than define this conversion. These mechanisms should therefore be tested comparatively by holding constant longitudinal exposure to permission-based sensing, behavioral tracking, synchronization, and AI inference while varying platform concentration, regulatory regime, substitutes, state-platform relations, and service bundling. Such studies could pair longitudinal user accounts with technical audits of data flows and platform behavior to examine when infrastructure exposure produces adaptation, resistance, or resignation. Several limitations bound the present claims. This qualitative study supports interpretive rather than causal generalization, captures users' situated accounts and folk models rather than verified system behavior, and reflects the research team's long-term relationship with participants. These limits define the contribution's scope: a China-based account of mechanisms that future UbiComp privacy research can compare, refine, and challenge.

\section{Conclusion}

This decade-long longitudinal study of 31 Chinese smartphone users reveals 
that digital resignation emerges through cumulative infrastructure exposure 
across three stages: functional acceptance of permissions, vigilance triggered 
by behavioral tracking, and eventual resignation as AI-driven inference made 
ecosystem-level profiling feel difficult to escape. Resignation was not 
indifference but exhaustion. Each new infrastructure layer retroactively 
reinterpreted earlier permissions and eroded users' capacity for 
self-management. These findings 
challenge permission-centric design assumptions and propose three system-level 
responses: inferential privacy visibility, context isolation, and automated privacy agency, 
shifting mobile privacy support from individual micro-management toward 
infrastructure-level intervention commensurate with the scale of modern mobile 
ecosystems.

\bibliographystyle{apalike}
\bibliography{reference}

\appendix

\section{Participant Characteristics}

\begingroup
\scriptsize
\setlength{\tabcolsep}{2.5pt}
\renewcommand{\arraystretch}{1.05}

\begin{xltabular}{\linewidth}{
>{\raggedright\arraybackslash}p{0.35cm}
>{\centering\arraybackslash}p{0.42cm}
>{\centering\arraybackslash}p{0.38cm}
>{\centering\arraybackslash}p{0.42cm}
>{\raggedright\arraybackslash}X
>{\raggedright\arraybackslash}X
>{\centering\arraybackslash}p{0.42cm}
>{\raggedright\arraybackslash}X
>{\raggedright\arraybackslash}X
>{\centering\arraybackslash}p{0.42cm}
>{\raggedright\arraybackslash}X
>{\raggedright\arraybackslash}X
>{\centering\arraybackslash}p{0.42cm}
>{\raggedright\arraybackslash}X
>{\raggedright\arraybackslash}X
}
\caption{Participant characteristics across the four interview waves. U/R indicates urban/rural residence; Edu. indicates educational attainment; Occ. indicates occupation. A slash (/) indicates missing data or non-participation in that wave.}
\label{tab:participant_characteristics}\\

\toprule
ID & Age & G. & 14 U/R & 14 Edu. & 14 Occ. & 17 U/R & 17 Edu. & 17 Occ. & 20 U/R & 20 Edu. & 20 Occ. & 23 U/R & 23 Edu. & 23 Occ. \\
\midrule
\endfirsthead

\toprule
ID & Age & G. & 14 U/R & 14 Edu. & 14 Occ. & 17 U/R & 17 Edu. & 17 Occ. & 20 U/R & 20 Edu. & 20 Occ. & 23 U/R & 23 Edu. & 23 Occ. \\
\midrule
\endhead

\midrule
\multicolumn{15}{r}{Continued on next page} \\
\midrule
\endfoot

\bottomrule
\endlastfoot

1  & 22 & F & R & High School        & Student        & U & Bachelor           & Student        & U & Master             & Student        & U & PhD                & Student        \\
2  & 31 & F & R & Bachelor           & Engineer       & R & Bachelor           & Engineer       & R & Bachelor           & NO             & R & Bachelor           & Engineer       \\
3  & 27 & M & R & Junior High School & Driver         & R & Junior High School & Driver         & R & Junior High School & Driver         & U & Junior High School & Driver         \\
4  & 28 & M & U & Bachelor           & Data Analyst   & U & Bachelor           & Data Analyst   & U & Bachelor           & Programmer     & U & Bachelor           & Programmer     \\
5  & 29 & F & U & Primary School     & Waiter         & U & Primary School     & Waiter         & U & Primary School     & Waiter         & U & Primary School     & Cleaner        \\
6  & 33 & M & R & Bachelor           & Programmer     & R & Bachelor           & Programmer     & R & Bachelor           & Programmer     & R & Bachelor           & Programmer     \\
7  & 35 & M & R & High School        & Chef           & R & High School        & Chef           & R & High School        & Chef           & R & High School        & Barber         \\
8  & 45 & F & U & Bachelor           & Journalist     & U & Bachelor           & Journalist     & U & Bachelor           & Journalist     & U & Bachelor           & Manager        \\
9  & 44 & F & R & High School        & Cashier        & R & High School        & Cashier        & R & High School        & Cashier        & R & High School        & Cashier        \\
10 & 36 & M & R & Bachelor           & Salesperson    & R & Bachelor           & Salesperson    & R & Bachelor           & Security Guard & R & Bachelor           & Designer       \\
11 & 38 & M & U & Junior High School & Farmer         & U & Junior High School & Farmer         & U & Junior High School & Farmer         & U & Junior High School & Farmer         \\
12 & 29 & F & U & Bachelor           & Laborer        & U & Bachelor           & Laborer        & U & Bachelor           & Laborer        & R & Bachelor           & Laborer        \\
13 & 25 & F & R & Bachelor           & Model          & R & Bachelor           & Model          & R & Bachelor           & Model          & R & Bachelor           & Model          \\
14 & 22 & M & R & High School        & Student        & R & Bachelor           & Student        & R & Master             & Writer         & R & Master             & Writer         \\
15 & 26 & M & R & Master             & Designer       & R & Master             & Designer       & R & Master             & Designer       & R & PhD                & Student        \\
16 & 27 & M & U & PhD                & Lecturer       & U & PhD                & Professor      & U & PhD                & Professor      & U & PhD                & Professor      \\
17 & 19 & M & U & High School        & Student        & U & Bachelor           & Student        & U & Bachelor           & Editor         & U & Master             & Student        \\
18 & 28 & F & U & Bachelor           & Teacher        & U & Bachelor           & Teacher        & U & Bachelor           & Teacher        & U & Bachelor           & Teacher        \\
19 & 34 & F & R & Primary School     & Janitor        & R & Primary School     & Janitor        & R & Primary School     & Janitor        & R & Primary School     & No             \\
20 & 35 & M & R & Bachelor           & Carpenter      & U & Bachelor           & Carpenter      & U & Bachelor           & Carpenter      & / & /                  & /              \\
21 & 37 & F & U & High School        & Carpenter      & U & High School        & Carpenter      & U & High School        & Gatekeeper     & / & /                  & /              \\
22 & 39 & M & R & Only literate      & Hairdresser    & R & Only literate      & Hairdresser    & R & Only literate      & Hairdresser    & R & Only literate      & Hairdresser    \\
23 & 47 & M & R & High School        & Factory Worker & R & High School        & Factory Worker & R & High School        & Factory Worker & R & High School        & Factory Worker \\
24 & 49 & F & U & Junior High School & Crew           & U & Junior High School & Crew           & / & /                  & /              & / & /                  & /              \\
25 & 52 & F & U & High School        & Carpenter      & U & High School        & Carpenter      & U & High School        & Cleaner        & U & High School        & No             \\
26 & 54 & F & R & High School        & Hairdresser    & R & High School        & Gatekeeper     & R & High School        & Gatekeeper     & R & High School        & Gatekeeper     \\
27 & 43 & M & R & Junior High School & Bartender      & / & /                  & /              & R & Junior High School & Bartender      & / & /                  & /              \\
28 & 44 & M & U & Bachelor           & Accountant     & U & Bachelor           & Accountant     & U & Bachelor           & Accountant     & R & Bachelor           & Accountant     \\
29 & 29 & F & U & Junior High School & Sex Worker     & U & Junior High School & Sex Worker     & U & Junior High School & Cleaner        & U & Junior High School & Cleaner        \\
30 & 26 & F & R & Bachelor           & Researcher     & R & Bachelor           & Researcher     & / & /                  & /              & / & /                  & /              \\
31 & 37 & M & U & Master             & Researcher     & / & /                  & /              & / & /                  & /              & / & /                  & /              \\

\end{xltabular}
\endgroup

\section{Interview Protocol}

\begin{longtable}{>{\raggedright\arraybackslash}p{3cm} >{\raggedright\arraybackslash}p{5.5cm} >{\raggedright\arraybackslash}p{6cm}}
\toprule
Section & Core Question & Possible Follow-up Probes \\
\midrule
\endfirsthead
\toprule
Section & Core Question & Possible Follow-up Probes \\
\midrule
\endhead
Smartphone Use & Could you describe how you usually use your smartphone in your everyday life? & Approximate daily usage time; typical situations in which you use your phone (e.g., at home, during commuting, or during work breaks); the role of smartphones in your daily routines. \\
Smartphone Use & What mobile apps or platforms do you currently use most frequently? & Main purposes of different apps; which apps are used most often; the role of different apps in your daily life. \\
Smartphone Use & In what situations do you typically use these apps? & Communicating with others, obtaining information, entertainment, work-related or study-related activities; whether there are regular usage patterns or routines. \\
Personal Information & When using mobile apps, what types of information do you think are related to you personally? & For example, account information, contact information, location, browsing history, or other types of information; which types of information you consider more important. \\
Privacy Understanding & Among these types of information, are there any that you consider more private or requiring special protection? & Why you consider certain information more private; whether different types of information have different levels of sensitivity for you. \\
Information Awareness & When using mobile apps, have you ever thought about whether these apps might record or store information related to you? & If so, in what situations did this thought arise; what kinds of information you think might be recorded. \\
Timeline Elicitation & Looking back at your experience since you first started using smartphones, have you noticed any changes in mobile apps, technologies, or your own usage practices over time? & You may recall different periods, such as when you first started using smartphones, several years later, and more recently; what kinds of changes you noticed and how you became aware of them. \\
Technology Experiences & During your use of mobile apps, have you encountered any features, system behaviors, or phenomena that left a strong impression on you? & When these features or phenomena appeared; why they were memorable; how you interpreted them at the time. \\
Critical Incident & Was there any particular experience that made you start paying closer attention to issues related to personal information when using mobile apps? & What happened in that situation; what made you become aware of the issue; your immediate reaction; whether this experience influenced your later usage practices. \\
App Permissions & When installing or using apps, have you noticed that some apps request access to certain permissions, such as location, contacts, photos, or microphone? & Whether you usually read these prompts; how you decide whether to allow these permissions; whether you have ever denied certain permissions. \\
Interpretation & When apps display certain features or behaviors, how do you usually interpret how these phenomena occur? & Whether they might be related to system functions, stored information, or other technological processes; how your understanding of these phenomena developed. \\
Information Use & How do you think mobile apps typically use or handle information related to their users? & For example, to provide functions, improve services, or other purposes; how you understand these processes. \\
Risk Perception & In what situations would you feel that certain apps or functions might involve issues related to personal information? & Whether particular contexts raise concerns; whether certain types of apps are more likely to create this feeling. \\
Behavioral Practices & In your daily smartphone use, have you taken any actions to reduce the amount of personal information that may be recorded or used? & For example, disabling certain permissions, limiting location access, sharing less information, or avoiding certain apps. \\
Behavioral Timeline & If you have taken such actions, when did you begin to do so? & Whether these practices started after a specific experience or gradually developed over time. \\
Practical Challenges & When trying to manage personal information on your smartphone, have you encountered any difficulties? & For example, complex settings, uncertainty about how to adjust settings, unclear prompts, or concerns about affecting app functionality. \\
Reflection & Looking back at your experiences using smartphones over the years, do you think your usage practices or perspectives have changed in any way? & Changes in trust toward apps, attention to personal information issues, or shifts in usage habits. \\
Closing Question & Is there anything else about smartphone use, app behaviors, or personal information that you would like to share? & Any experiences, observations, or thoughts that you think are important but have not yet been mentioned. \\
\bottomrule
\end{longtable}

\section{Analytic Procedure Details}

The research team classified each participant's long term trajectory of privacy understanding based on their overall narrative across multiple interview waves, considering participants' dominant interpretive framework in each wave and how that framework changed over time. The research team first read each participant's complete longitudinal interview record, then compared statements across different time points. A participant was judged to be in a given stage only when the core features of that stage were consistently supported across multiple points in the corresponding wave and had become the dominant interpretive framework for that period; isolated or occasional related statements that did not constitute the participant's primary framework were not treated as evidence of a stage transition.

Stage 1 was defined by participants primarily understanding data collection as a necessary cost of obtaining smartphone functions and services, with limited attention to how personal data were subsequently used. Their accounts typically centered on convenience, functional availability, and everyday use experience, with little spontaneous discussion of data flows, privacy risk, or long term consequences.

Stage 2 was defined by participants explicitly recognizing that platforms continuously collected, analyzed, and made use of their behavioral data, and being able to explain this tracking mechanism with reference to specific experiences. For example, participants began to mention targeted advertising, cross application recommendations, browsing records, location data, or permission invocations, and gradually came to recognize that these forms of data were linked to one another.

Stage 3 was defined by participants explicitly identifying a further complication of privacy risk introduced by newer technologies, for example the capacity of AI systems to integrate data from multiple sources for cross context inference. At the same time, participants explicitly judged that conventional strategies such as permission management, privacy settings, or reduced use had become largely ineffective, and accordingly showed a shift toward recalibrated privacy expectations, limited acceptance, or an abandonment of further control.

A participant was judged to have undergone the complete three stage evolution when the core features of the three stages appeared successively across different interview waves, with each later stage gradually becoming the participant's dominant interpretive framework. Conversely, participants who remained within a single stage over an extended period, or who showed a transition between only two of the three stages, were classified as having undergone a single stage or a two stage evolution rather than the complete three stage trajectory. Agreement among the three researchers regarding which participants, among those who completed all interview waves, underwent the complete three stage evolution was high overall. Only the classification of P6 involved some disagreement, owing to a degree of internal inconsistency in P6's statements, which the research team presented as such in the reporting of findings; the judgment that P6 did not undergo the complete three stage evolution was nonetheless endorsed by all three researchers.

Disagreements encountered during open coding and the subsequent construction of thematic categories were addressed through discussion rather than majority vote. Such disagreements were concentrated in a small number of boundary cases, for instance, whether a particular statement should be classified under Stage 1 or Stage 2. In these instances, the research team returned repeatedly to the original transcripts and their surrounding context to review and discuss the material until a shared understanding was reached. This iterative process also contributed to the ongoing refinement of the analytical categories and the longitudinal analytic framework. When the three researchers were unable to reach agreement, the last author of this paper served as an arbitrator, and the last author's judgment served as the final decision; such instances occurred rarely.

Because participants sometimes revisited earlier experiences in later interviews, retrospective accounts did not always fully align with the corresponding contemporaneous transcript. When such inconsistencies arose, the research team treated each wave's transcript as the primary evidence of the participant's view at that time. Subsequent retrospective accounts were treated as reflecting participants' current interpretation of their past experiences, rather than as a replacement for the original record. The analysis did not attempt to resolve all such inconsistencies into a single coherent narrative; instead, these points of tension were retained, as they helped illuminate participants' evolving understanding. Longitudinal interpretation was formed by comparing records across different interview waves rather than by relying solely on retrospective reconstruction. In practice, instances of substantive contradiction in participants' statements were relatively infrequent in this study; most participants offered relatively clear accounts of the reasons underlying changes in behavior or attitude, or clear revisions of their earlier views.

The longitudinal design of this study required maintaining contact with participants across multiple interview waves, which helped establish trust and allowed participants to discuss experiences that might not otherwise surface in a single interview. At the same time, the research team recognized that this sustained relationship could also shape participants' accounts. Participants might become increasingly familiar with the research topic over repeated interviews, or might retrospectively organize their experiences into a more coherent developmental narrative than the experiences themselves warranted. To reduce this risk, the questions in each wave focused primarily on participants' current experiences, and the research team sought to avoid prompting participants to assess or anticipate changes the research team might expect. In the analysis, interpretation relied primarily on the contemporaneous record from each wave, while retrospective reflections were treated as supplementary data reflecting how participants currently understood their past experiences, rather than as direct evidence of their earlier views. Even with these measures in place, the research team recognizes that this study remains subject to the possible influence of the researchers' own position; this is noted further in the limitations section.

The role of researchers. \minoredit{Coding} and longitudinal analysis were conducted by three researchers (the first three authors), all with backgrounds in HCI and ubiquitous computing, who participated continuously throughout the study with no personnel change. The team member responsible for maintaining contact with participants between waves (see Section 3.2) was not involved in coding, so relationship-building and interpretive judgment did not overlap. The last author coordinated the overall analytic process, reviewed the coding results, and arbitrated disagreements among the three coders. None of the researchers held any employment or financial relationship with the platforms discussed by participants.

Reflections on researcher positionality. This approach carries two limitations. First, the shared HCI and ubiquitous computing training of the three coders may have shaped how analytic categories were constructed (for example, framing privacy around interactional mechanisms), in ways that researchers from other disciplinary backgrounds might not have. Second, sustained contact with participants may have increased the team's familiarity with their earlier accounts over time; we mitigated this by keeping each wave's interview questions focused on participants' current experiences and by treating the contemporaneous transcript, rather than later retrospective accounts, as the primary evidence.

\section{\minoredit{Additional Cross-Stage and Supporting Quotations from Participants}}
\label{app:additional_quotes}

\begin{table}[htbp]
\caption{Cross stage comparison of representative participant quotations (2014, 2017, 2020, 2023)}
\label{tab:appendix_quotes}
\centering
\footnotesize
\setlength{\tabcolsep}{3pt}
\renewcommand{\arraystretch}{1.1}

\begin{tabular}{
p{0.8cm}
p{3.8cm}
p{3.8cm}
p{3.8cm}
p{3.8cm}
}
\toprule
\textbf{ID} &
\textbf{2014} &
\textbf{2017} &
\textbf{2020} &
\textbf{2023} \\
\midrule

P5 &
Obviously I do not think it is a big problem for camera data to be used. &
I realized how much information the phone was taking, and what that information could be used for. &
I now use two accounts. &
I am really tired. \\
\addlinespace

P7 &
Many apps ask for location permission during registration, like maps do. I thought that was reasonable, so I agreed. &
Maps might be accessing too much data, some of which is probably unnecessary. &
There are many strategies you can use. For example, blocking voice access alone makes things a lot safer. &
I feel that compared with all the information AI already collects about me, one less voice message really makes no difference. \\
\addlinespace

P12 &
I do not really mind. Is not this data used to help with verification and judgment anyway? &
Whenever that indicator light comes on and tells me something is being recorded, I always feel a bit tense and uncomfortable. &
I can proactively turn off things like location, and also visit fewer questionable and low quality websites. Both help. &
But AI, and modern technology in general, has become so powerful that it is very hard to respond to it. \\
\addlinespace

P15 &
I thought that was fine. If it does not know my contacts, how could it recommend people to me? &
Something is wrong. Something is definitely wrong. &
They must be collecting my data and using it, and I had no idea at all. &
It does not matter anymore. It can be however it wants to be. \\
\addlinespace

P16 &
There is no need to worry about privacy issues related to smart data collection. &
If the data is fairly sensitive, I delete all of it, since that feels safer. &
Sometimes I am actually grateful that it keeps reminding me it is recording. &
I increasingly feel that I am not talking to an AI on my phone, but to a person on the other side of the screen who can infer things about me from the smallest details. \\
\addlinespace

P19 &
I think the functions we are discussing here are all normal. They naturally need this data in order to work. &
Once you have searched for something, it will keep recommending it to you afterward. &
Now I proactively avoid using certain apps on my phone, especially malicious programs. &
I am really tired. \\
\addlinespace

P23 &
You use a camera to take photos, and of course a camera needs to access the camera hardware. Otherwise how could it take photos? &
It is necessary to avoid capturing background details and other unrelated information in a photo, since it may contain private information we ourselves are not even aware of. &
Some strategies do not really work, but overall I feel things are a lot safer. &
You can say less, but as for how much it actually records, you really do not know. \\
\addlinespace

P26 &
If you buy something, of course they need your location. Just turn it on, what privacy problem could that possibly cause? &
Many apps are often malicious, collecting more information than they should. &
Many of the measures I take are actually ineffective, and that frustrates me, but I am still taking action. &
Honestly, I cannot even find a good solution to this, meaning the privacy problem. \\

\bottomrule
\end{tabular}
\end{table}

\subsection*{\minoredit{Additional supporting quotations from Sections 4.2 and 4.3}}

\minoredit{The following quotations provide supplementary examples of mechanisms illustrated in Sections 4.2 and 4.3.}

\paragraph{\minoredit{Behavioral feedback (Section 4.2).}}
\minoredit{\textbf{P14 (2020):} \textit{``Sometimes I feel my phone understands me to a scary degree. I was still debating what Chinese food I wanted that day, and then I opened the delivery app and every single recommendation was exactly what I wanted.''}}

\paragraph{\minoredit{System visibility (Section 4.2).}}
\minoredit{\textbf{P16 (2020):} \textit{``Sometimes I'm actually grateful that it keeps reminding me when it's recording. In one app, they visualized it with a headphone icon. That made me much more sensitive to it.''}}

\paragraph{\minoredit{Reduced public expression (Section 4.2).}}
\minoredit{\textbf{P14 (2020):} \textit{``I hardly post life updates anymore. It doesn't feel necessary, and every time I do, I worry about what unexpected things other people might infer from that information.''}}

\paragraph{\minoredit{Concern about AI (Section 4.3).}}
\minoredit{\textbf{P12 (2023):} \textit{``How could this, meaning privacy during phone use, not be important? But AI, and modern technology in general, has become so powerful that it is very hard to respond to it.''}}

\paragraph{\minoredit{Residual protective measures (Section 4.3).}}
\minoredit{\textbf{P17 (2023):} \textit{``I won't reveal identity information, I know how important identity is. Other than that, there isn't really much else. Everyone is basically running around online with nothing to hide behind anyway. The technology is just too powerful, so what difference does it make.''}}

\clearpage
\noindent
\begin{minipage}{\textwidth}
\section{Portability of Findings Across Structural Conditions}
\label{app:portability}

\small
\setlength{\tabcolsep}{4pt}
\captionof{table}{Extent to which key findings are shaped by China-specific structural features.}
\label{tab:portability}
\begin{tabular}{p{4.6cm}p{1.3cm}p{2.3cm}p{6.0cm}}
\toprule
\textbf{Finding} & \textbf{Location} & \textbf{Affected by China-specific structure?} & \textbf{China-specific factor (if applicable)} \\
\midrule
Functional framing of permission requests (contacts, photos, camera, microphone, location) as necessary for task completion & \minoredit{4.1} & No clear effect observed & --- \\
\addlinespace
Behavioral feedback (repeated ads, recommendation reinforcement) triggering suspicion of background tracking & \minoredit{4.2} & No clear effect observed & --- \\
\addlinespace
System visibility (microphone/camera indicators, access logs) prompting reassessment of prior permissions & \minoredit{4.2} & No clear effect observed & --- \\
\addlinespace
Cross-platform advertising echoes (e.g., browsing on one app surfacing recommendations on unrelated apps) & \minoredit{4.2} & Partially affected & This pattern can occur wherever apps share data through ad networks or third-party SDKs; it may be more readily noticed where app-market concentration and cross-app data sharing are more common, as in China \\
\addlinespace
Limited-adjustment coping strategies (account separation, guest browsing, deleting search history) & \minoredit{4.2, 4.3} & No clear effect observed & --- \\
\addlinespace
AI-driven multi-source inference producing a felt ``complete picture'' of the user & \minoredit{4.3} & Strongly affected & Requires voice, chat, location, and device data to sit within one platform/account ecosystem, as in China's super-app architecture~\cite{jia2022superapps} \\
\addlinespace
Sense of being unable to opt out of data flows entirely (P11) & \minoredit{4.3} & Strongly affected & Bundling of payment, social, mobility, and identity services within a small number of super apps limits exit options~\cite{jia2022superapps,plantin2019wechat} \\
\addlinespace
Cognitive overload from managing permissions, tracking, and AI inference simultaneously & \minoredit{4.3} & Partially affected & The co-occurrence of all three layers is common to modern mobile ecosystems generally; the burden may be felt more acutely where these functions are concentrated on a single platform~\cite{jia2022superapps} \\
\addlinespace
Retrospective reinterpretation of earlier permissions after later infrastructure becomes visible & \minoredit{4.2} & No clear effect observed & --- \\
\bottomrule
\end{tabular}
\end{minipage}

\end{document}